\documentclass[apj]{emulateapj}

\def\ks{km~s$^{-1}$}
\def\ms{m~s$^{-1}$}

\def\mjup{M$_{\rm Jup}$}

\def\msun{M$_{\odot}$}
\def\rsun{R$_{\odot}$}
\def\lsun{R$_{\odot}$}
\def\msini{$M_P\sin i~$}
\def\chisq{$\sqrt{\chi^2_\nu}$}
\def\chis{$\chi^2_\nu$}
\def\plmn{~$\pm$~}
\def\feh{[Fe/H]}

\def\logg{$\log{g}$}

\def\vsini{$V_{\rm rot}\sin{i}$}

\def\starA{HD\,102956}
\def\pA{6.4950}
\def\peA{0.0004}

\def\tpA{2455346}
\def\tpeA{0.7}
\def\eA{0.048}
\def\eeA{0.027}

\def\omA{12}
\def\omeA{40}
\def\kA{73.7}
\def\keA{1.9}
\def\msiniA{0.96}
\def\msinieA{0.05}
\def\arelA{0.081}
\def\areleA{0.002}
\def\rmsA{6.0}
\def\chisA{1.35}
\def\nobsA{22}
\def\mstarA{1.68}
\def\mstareA{0.11}
\def\bvA{0.971}
\def\bveA{0.01}
\def\vmagA{8.02}
\def\vmageA{0.05}
\def\mvA{2.5}
\def\mveA{0.2}
\def\vsiniA{0.30}
\def\ageA{2.3}
\def\ageeA{0.5}
\def\rstarA{4.4}
\def\rstareA{0.1}
\def\lstarA{11.6}
\def\lstareA{0.5}
\def\teffA{5054}
\def\loggA{3.5}
\def\feA{+0.19}

\def\dA{126}
\def\deA{13}

\def\sval{0.17}
\def\rhk{-5.09}
\def\starA{HD\,102956}

\begin{document}
\title{A hot Jupiter orbiting the 1.7 \msun\ Subgiant \starA$^1$}  

\author{John Asher Johnson\altaffilmark{2},
Brendan P. Bowler\altaffilmark{3},
Andrew W. Howard\altaffilmark{4},
Gregory W. Henry\altaffilmark{5},
Geoffrey W. Marcy\altaffilmark{4},
Howard Isaacson\altaffilmark{4},
John Michael Brewer\altaffilmark{6},
Debra A. Fischer\altaffilmark{6},
Timothy D. Morton\altaffilmark{2},
Justin R. Crepp\altaffilmark{2}
}

\email{johnjohn@astro.berkeley.edu}

\altaffiltext{1}{Based on observations obtained at the
W.M. Keck Observatory, which is operated jointly by the
University of California and the California Institute of
Technology. Keck time has been granted by both NASA and
the University of California.}
\altaffiltext{2}{Department of Astrophysics,
  California Institute of Technology, MC 249-17, Pasadena, CA 91125;
  NASA Exoplanet Science Institute (NExScI)}
\altaffiltext{3}{Institute for Astronomy, University of Hawai'i, 2680
  Woodlawn Drive, Honolulu, HI 96822} 
\altaffiltext{4}{Department of Astronomy, University of California,
  Mail Code 3411, Berkeley, CA 94720}
\altaffiltext{5}{Center of Excellence in Information Systems, Tennessee
  State University, 3500 John A. Merritt Blvd., Box 9501, Nashville, TN 37209}
\altaffiltext{6}{Department of Astronomy, Yale University, New Haven,  CT 06511}

\begin{abstract}
We report the detection of a giant planet in
a \pA~day orbit around the \mstarA~\msun\ subgiant \starA. The planet has
a semimajor axis $a = \arelA$~AU and minimum mass \msini~$ =
\msiniA$~\mjup. \starA\ is the most massive star known to harbor a hot
Jupiter, and its planet is only the third known to orbit within 0.6~AU
of a star more massive than 1.5~\msun. Based on our sample of 137
subgiants with $M_\star > 1.45$~\msun\ we find that 0.5--2.3\% of A-type
stars harbor a close-in planet ($a < 0.1$~AU) with \msini $>$ 1 \mjup,
consistent with hot-Jupiter occurrence for Sun-like stars. 
Thus, the paucity of planets with $0.1 < a < 1.0$~AU around
intermediate-mass stars 
may be an exaggerated version of the ``period valley'' that is
characteristic of planets around Sun-like stars. 
\end{abstract}

\keywords{techniques: radial velocities---stars: individual (\starA)---planets and satellites: formation}

\section{Introduction}

\begin{figure}[!t]
\epsscale{1.1}
\plotone{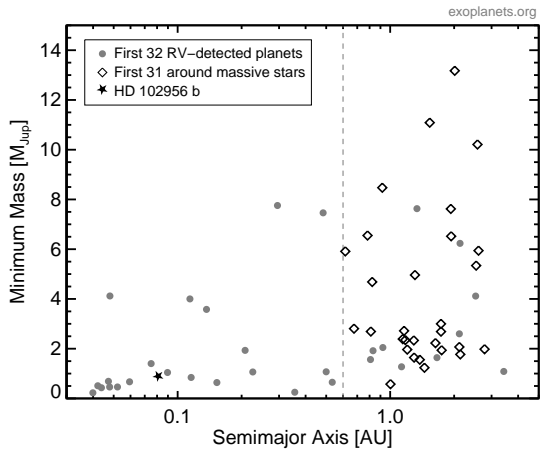}
\caption{The semimajor axes and minimum masses (\msini) of the first
  32 Doppler-detected planets (gray circles). The majority of these
  planets orbit stars with $M_\star < 1.5$~\msun, and the planets span
  a wide range of semimajor axes. Also shown are the first 31
  Doppler-detected planets around massive stars ($M_\star >
  1.5$~\msun; open diamonds), and the sole hot Jupiter, \starA\,b
  (filled five-point star). Compared to the population of planets
  around Sun-like  stars, there is a notable paucity of planets 
  inward of 0.6~AU around the intermediate-mass
  stars.  \label{fig:firstplanets}}  
\end{figure}

The current state of knowledge of planets around intermediate-mass
($M_\star \gtrsim 1.5$~\msun) stars is reminiscent of the general knowledge of
exoplanets in 2001. At that time there were 32 planets known, mostly
orbiting Sun-like stars. The distributions of semimajor axes and
masses (\msini) of these early exoplanet discoveries, drawn from the
Exoplanet Orbit 
Database\footnote{\tt{http://exoplanets.org}}, are shown in
Figure~\ref{fig:firstplanets}, illustrating the
prevalence of giant planets in unexpectedly close-in
orbits\footnote{While unexpected 
  based on the sample of one provided by the Solar System, the
  existence of short-period Jovian planets was predicted in at least
  one published instance prior to the discovery of the first hot
  Jupiter \citep{mayor95}. In 1952 Otto Struve
  mused,  ``...[T]here seems to be no compelling reason 
  why the...planets should not, in some instances,
  be much closer to their parent star than is the case in the solar
  system. It would be of interest to test whether there are any such
  objects.'' \citep{struve52}}. As of 2010 May there are 31 planets
known to orbit intermediate-mass stars ($M_\star > 1.5$~\msun), and the
semimajor axes of planets in this new class was surprising, but for a
different reason: 
there are no planets orbiting closer than 0.6~AU
\citep{johnson07,sato08a}. As \citet{bowler10} showed,
the  planet populations orbiting host stars on either side of
1.5~\msun\ are distinct at the 4-$\sigma$ level.  

Close-in, Jovian planets are 
relatively easy to detect using radial velocities because of the larger
amplitude they induce and the increased number of orbit cycles per
observing time baseline. Their absence therefore cannot be due to an
observational bias given the large population of planets at longer
periods. Instead, it appears that 
stellar mass has a dramatic effect on the
semimajor axis distribution of planets. However, it is not clear
whether this effect is a reflection of the process of planet formation
and migration, or instead related to the effects of the evolution of
the host stars.   

Stellar evolution may be an important factor because Doppler surveys 
of intermediate-mass stars are largely restricted to
post-main-sequence targets. While massive,
main-sequence stars are poor Doppler targets due to their rapid
rotation (\vsini~$\gtrsim 50$~\ks; \citet{lagrange09a}), their evolved
counterparts on the giant and subgiant branches are much slower
rotators (\vsini~$\lesssim 5$~\ks) and therefore have the narrow
absorption lines required for precise Doppler measurements. However,
as stars evolve their atmospheres expand and may encroach upon the
orbits of their planets. Simulations by \citet{nordhaus10}, 
\citet{carlberg09} and \citet{villaver09} have
suggested that the engulfment planets by the expanding
atmospheres of stars can account for the lack of close-in planets
around K-giants and clump-giants \citep[see
  also][]{sato08a}.  

While Doppler surveys have encountered a barren region around A stars
inward of 0.6~AU, transit surveys have discovered two examples of hot
Jupiters around intermediate-mass stars. OGLE2-TR-L9 and WASP-33 are
1.5~\msun\ stars orbited by Jovian planets with semimajor axes $a =
0.041$~AU and 0.026~AU, respectively \citep{snellen09,
  cameron10}. These detections demonstrate 
that close-in planets exist around A-type dwarfs,
adding additional concern that the lack of planets close to evolved 
intermediate-mass stars is the result of stellar
engulfment. Unfortunately, the complicated observational and selection
biases inherent to ground-based, wide-field transit surveys make it
difficult to 
measure accurate occurrence rates that can be meaningfully compared
to those measured from Doppler surveys \citep{gaudi05}. 

Among the various types of evolved stars, subgiants offer a unique
view of the population of hot Jupiters around intermediate-mass
stars. The radii of subgiants have inflated by only a factor of $\approx 
2$ compared to their main-sequence values. The simulations of
\citet{villaver09} show that planets with $a \gtrsim 0.1$~AU
should be safe from the tidal influence of stars near the base of the
RGB (see their Figure 2). It is only after stars start to ascend the
RGB and have their radii expand to an appreciable fraction of an AU
that tidal influences become important. The occurrence rates and
semimajor axis distribution of close-in planets around subgiants should
therefore be representative of the properties of planets around
A-type dwarfs. 

We are conducting a Doppler survey of intermediate-mass subgiants at
Keck and Lick Observatories to study the effects of stellar mass on
the physical properties and orbital architectures of planetary
systems. Our survey has resulted in the detection 
of 14 planets around 12 intermediate-mass ($M_\star \gtrsim
1.5$~\msun) stars, and 4 additional planets around less-massive
subgiants \citep[][Johnson et al. 2010b, submitted]{johnson06b,
  johnson07,johnson08b,bowler10, peek09, johnson10b}. In this Letter
we report the first Doppler-detected 
planet within 0.6~AU of a ``retired'' (former) A star: a hot Jupiter
around a \mstarA~\msun\ subgiant.   

\section{Stellar Properties, Radial Velocities and Orbit}

\starA\ (=HIP\,57820) is listed in the \emph{Hipparcos} Catalog with 
$V=\vmagA$, $B-V = \bvA$, a 
parallax--based distance of \dA~pc, and an absolute 
magnitude $M_V = \mvA$ \citep{hipp}.  Like most subgiants, 
\starA\ is chromospherically--quiet with an average $S = \sval \pm
0.02$ and  $\log{R^\prime_{HK}} = \rhk$ on the Mt. Wilson scale
\citep{wright04b}. We used the LTE spectral synthesis 
described by \citet{valenti05} and \citet{fischer05b} to estimate the
spectroscopic properties. To constrain the low surface
gravities of the evolved stars we used the iterative scheme of
\citet{valenti09}, which ties the SME-derived value of $\log{g}$ to
the gravity inferred from the Yonsei-Yale  (Y$^2$) stellar models. Our
SME analysis gives  $T_{\rm eff} = \teffA \pm 44$~K,
$\rm [Fe/H] = \feA \pm 0.04$,
$\log{g} = \loggA \pm 0.06$ and
$V_{\rm rot}\sin{i} = \vsiniA \pm 0.5$~\ks. We compared the star's
temperature, luminosity and metallicity to the Y$^2$ stellar model
grids to estimate a stellar mass 
$M_\star = \mstarA \pm \mstareA$~\msun\ and radius $R_\star =
\rstarA$~\plmn0.07~\rsun. All of the stellar properties are summarized 
in Table \ref{tab:starorbit}.  

\begin{figure}[!t]
\epsscale{1.1}
\plotone{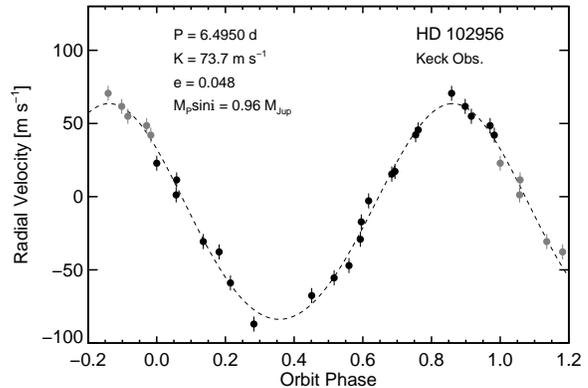}
\caption{Doppler measurements of \starA\ from Keck Observatory phased
  at the orbital period of the planet. The error
  bars are the quadrature sum of the internal measurement uncertainties
  and 5~\ms\ of jitter. The dashed line shows the best-fitting orbit
  solution of a single Keplerian orbit. The solution results in
  residuals with an rms scatter of \rmsA~\ms\ and
  \chisq~$=$~\chisA, indicating a good fit to the data. \label{fig:orbit}}
\end{figure}

We began monitoring the radial velocity of \starA\ in 2007 April and
we have gathered a total of \nobsA\ measurements. After two seasons of
observing we noticed RV variability with an rms scatter of 41~\ms,
which is much larger than the scatter predicted by the measurement
uncertainties and jitter levels typical of subgiants
\citep{fischer03, wright05, johnson10b}. 
Table~\ref{tab:rv} lists our RV measurements, times of observation and
internal errors (without jitter). \citet{johnson10b} estimate a
typical jitter of 5~\ms\ based on their analysis of 382
RV observations of 72 stable 
subgiants observed at Keck with HIRES. We add this jitter 
estimate in quadrature to the internal measurement errors to ensure
proper weighting of the data in our orbit analysis. 

To search for the best-fitting orbit we used the partially-linearized
Keplerian fitting code 
\texttt{RVLIN}\footnote{http://exoplanets.org/code/} 
described by \citet{wrighthoward}. As an alternative to a periodogram
analysis, we first stepped through a grid of orbital
periods sampled from 1 to 100 days in 5000 equal,
logarithmically-spaced intervals. At each step we fixed  
the period, searched for the best-fitting orbit and recorded the
resulting\footnote{We use \chisq\ to indicate the factor by which
the observed scatter about the best-fitting model differs from our
expectation based on the measurement errors. Thus, the 
scatter about our model is a factor of \chisA\ larger than our
average error bar.} \chisq. We found a minimum (\chisq~$ = 1.15$) at periods near
6.5~days. The next 
lowest minimum (\chisq~$ = 1.98$) occurs at $P = 3.47$~days. However,
with a best-fitting eccentricity of 0.95, the orbit solution is clearly
unphysical.

We then allowed the period to float in our {\tt RVLIN} analysis with
an initial guess of $P= 6.5$~days, and found that a single-planet
Keplerian model with a period $P = 
\pA \pm \peA$~days, eccentricity $e = \eA \pm \eeA$ and velocity semiamplitude
$K = \kA \pm \keA$~\ms. The fit produces RV residuals with a
root-mean-squared (rms) scatter of \rmsA~\ms\ and reduced \chisq~$ =
\chisA$, indicating an 
acceptable fit. We used the false-alarm analysis of
\citet{howard09} to calculate FAP~$ < 0.001$ \citep[see
  also][]{johnson10a}. The resulting minimum  
planet mass is \msini~$=\msiniA$~\mjup, and the semimajor axis is $a =
\arelA$~AU. The best-fitting solution is shown in Figure~\ref{fig:orbit},
where the plotted error bars are the quadrature sum of internal errors and
5~\ms\ of jitter.

\begin{deluxetable}{lc}
\tablecaption{Stellar Properties and Orbital Solution for \starA
\label{tab:starorbit}}
\tablewidth{0pt}
\tablehead{
\colhead{Parameter} & \colhead{Value} 
}
\startdata
$V$   & $\vmagA \pm \vmageA$ \\
$B-V$ & $\bvA \pm \bveA$ \\
$Distance$ (pc) & $\dA \pm \deA$ \\
$M_V$ & $\mvA \pm \mveA$ \\
\feh & $\feA \pm 0.04$ \\
T$_{\rm eff}$ (K) & $\teffA \pm 44$ \\
\vsini~(\ks) & $\vsiniA \pm 0.5$ \\
\logg    & $\loggA \pm 0.06$ \\
$M_\star$ (\msun) & $\mstarA \pm \mstareA$ \\
$R_\star$ (\rsun) & $\rstarA \pm \rstareA$ \\
$L_\star$ (\lsun) & $\lstarA \pm \lstareA$ \\
$\log R'_{HK}$  & $\rhk$ \\
Age (Gyr) & $\ageA \pm \ageeA$ \\
 &  \\
$P$ (days) &  $\pA \pm \peA$ \\
$K$ (m\,s$^{-1}$) & $\kA \pm \keA$ \\
$e$ & $\eA \pm \eeA$\\
$T_P$ (Julian Date) & $\tpA \pm \tpeA$ \\
$\omega$ (degrees) & $\omA \pm \omeA$ \\
\msini\ (\mjup) & $\msiniA \pm \msinieA$ \\
$a$ (AU) & $\arelA \pm \areleA$ \\ 
$R_\star/a$ & $0.253 \pm 0.008$ \\
N$_{\rm obs}$ & \nobsA \\
rms (m\,s$^{-1}$) & \rmsA \\
\chisq & \chisA \\
\enddata
\end{deluxetable}

After identifying the best-fitting model, we use a Markov-Chain Monte
Carlo (MCMC) algorithm to 
estimate the parameter uncertainties \citep[See,
  e.g.][]{ford05,winn07,bowler10}.  MCMC is a Bayesian inference 
technique that uses the data to explore the shape of the likelihood
function for each parameter of an input model.  At each step, one 
parameter is selected at random and altered by drawing a random variate
from a normal distribution. If the resulting $\chi^2$ (not reduced by
the number of free parameters $\nu$) value for the new
trial orbit is less than the previous $\chi^2$ value, then the trial
orbital parameters are added to the chain.  If not, then the
probability of adopting the new value is set by the difference in
$\chi^2$ from the previous and current trial steps.  If the 
current trial is
rejected then the parameters from the previous step are adopted.  
We adjusted the width of the normal distributions from which the steps
are drawn until we achieved a 30--40\% acceptance rate in each parameter.
The resulting ``chains'' of parameters form the posterior probability
distribution, from which we select the 15.9 and 84.1
percentile levels in the cumulative distributions as the ``one-sigma''
confidence limits. In most cases the posterior probability
distributions were approximately Gaussian. The orbital
parameters and their uncertainties are listed in
Table~\ref{tab:starorbit}.  

As an additional check on the nature of the RV variations, we acquired
photometric observations of HD~102956 with the T3 0.4~m  automatic
photometric telescope (APT) at Fairborn Observatory.  A brief
description of the T3 data acquisition and reduction procedures, as
well  as the utility of the APT observations for eliminating false
positive detections, can be found in \citet{johnson10b}.

The APT collected two dozen observations in the Johnson $V$ and $B$ 
photometric bands between 2010 May 30 and June 23, near the end of the 
2010 observing season.  The $V$ observations scatter about their mean with 
a standard deviation of 0.0037 mag, consistent with T3's measurement 
precision for a single observation \citep[e.g.,][Tables 2--3]{henry00c}.  
A least-squares sine fit on the 6.4948-day radial velocity period yielded 
a semi-amplitude of $0.0012\pm0.0010$ mag.  Identical results
were obtained for the $B$ observations. This tight upper limit to 
photometric variability provides strong support for the planetary
interpretation of the radial velocity variations.

\section{The fraction of intermediate-mass stars with hot Jupiters}

Our Keck and Lick Doppler surveys of subgiant stars have time
baselines of 3 and 6 years, respectively, and the majority of the
target stars have more than 6 observations with jitter-limited
measurement precision ranging from 3--6~\ms. Our precision, cadence
and time baseline provide us with the opportunity to assess the
fraction of intermediate-mass stars with hot Jupiters, which we define as
planets with $a < 0.1$~AU. This definition is somewhat arbitrary, but
it is consistent with definitions widely used by other studies of hot
Jupiters, which typically focus on Solar-mass stars and periods $P
\lesssim 10$~days. The most comprehensive study of this kind is that
of \citet{cumming08}, who measure an occurrence rate of $f = 0.004 \pm
0.003$ for $a < 0.1$~AU, \msini~$\geq 1$~\mjup\ and primarily stars
with masses $M_\star < 1.4$~\msun. Their reported planet occurrence is
consistent with $f < 0.01$ at 95\% confidence.  

Within our sample we restrict our analysis to stars with $M_\star >
1.45$~\msun\ (the evolved counterparts of A-type stars), $N_{\rm
  obs} > 3$, \vsini~$ < 20$~\ks\ and no evidence of double-lines
indicative of an SB2. Our sample
contains 137 stars that meet these 
criteria. For each star we first perform a periodogram analysis and
use {\tt RVLIN} to search for orbit solutions near the strongest
periodicities using the same technique described by \citet{marcy05b}. 
We then evaluate the FAP by estimating the
likelihood of improving \chis\ over that of a linear fit
\citep[see][for further details]{howard09}. For solutions with FAP~$ <
0.01$, we record the rms of the residuals about the best-fitting
orbit. For larger FAP values we record the rms about the best-fitting
linear fit to the RVs. 

\begin{deluxetable}{lll}
\tablecaption{Radial Velocities for HD 102956\label{tab:rv}}
\tablewidth{0pt}
\tablehead{
\colhead{JD} &
\colhead{RV} &
\colhead{Uncertainty} \\
\colhead{-2440000} &
\colhead{(m~s$^{-1}$)} &
\colhead{(m~s$^{-1}$)} 
}
\startdata
14216.805 &   11.40 &  1.18 \\
14429.150 &    5.77 &  1.39 \\
14866.114 &   37.10 &  1.30 \\
14986.851 &  -58.52 &  1.27 \\
15014.772 &   59.14 &  1.07 \\
15016.870 &  -49.20 &  1.13 \\
15284.917 &  -79.04 &  1.34 \\
15285.993 &  -14.32 &  1.25 \\
15313.915 &   43.48 &  1.25 \\
15314.828 &  -10.25 &  1.18 \\
15343.795 &  -66.93 &  1.17 \\
15344.885 &    3.98 &  1.35 \\
15350.786 &  -40.57 &  1.27 \\
15351.880 &   34.14 &  1.19 \\
15372.804 &   30.58 &  1.16 \\
15373.796 &  -42.13 &  1.22 \\
15374.753 &  -98.50 &  1.10 \\
15376.783 &  -28.67 &  1.12 \\
15377.812 &   30.78 &  1.10 \\
15378.749 &   50.22 &  1.12 \\
15379.787 &    0.00 &  1.13 \\
15380.805 &  -70.40 &  1.13
\\
\enddata
\end{deluxetable}

Next, we measure the largest velocity semiamplitude $K_{\rm up}$
that is consistent with the observed rms scatter for simulated planets
of various periods. Our method is 
similar to that of \citet{lagrange09a} and \citet{bowler10}. For each
star we sample a range of orbital periods corresponding 
to semimajor axes $0.04 \leq a \leq 0.10$~AU\footnote{The lower limit of
$0.04$~AU corresponds to  $a/R_\star = 2$ for a typical $M_\star =
1.7$~\msun\ subgiant with $R_\star = 4$~\rsun, which is the smallest
scaled semimajor axis among the known hot Jupiters listed in the
Exoplanet Orbit Database.}. At each fixed period we generate a
sample of 3000 simulated orbits  
with random phases and circular orbits sampled at the actual times of
observation. For each simulated orbit we also add 5~\ms\ of random
noise to simulate jitter. We then record the 
distribution of 3000 simulated velocity rms values and compare the
distribution to the rms of the measurements. Finally, we adjust $K$
until the measured rms is less than that of 99.7\% of the simulated
orbits, and record the semiamplitude as $K_{\rm up}$ at that period.

\begin{figure}[!t]
\epsscale{1.2}
\plotone{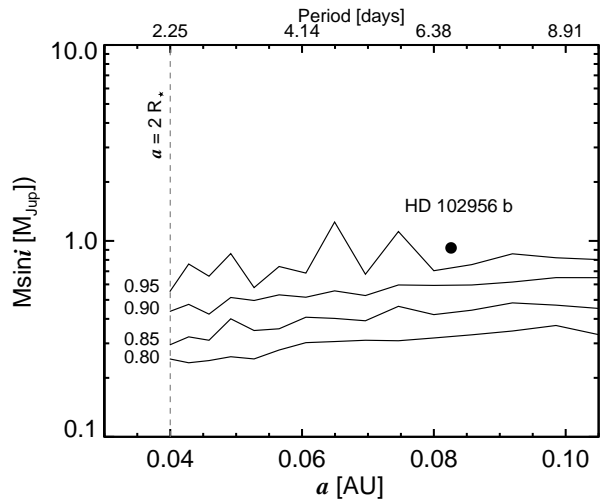}
\caption{Detection limits of the Keck Doppler survey of
  intermediate-mass subgiants as a function of semimajor axis. The
  contours indicate the minimum 
  planet mass detectable with a given fractional completeness, which
  is labeled to the left of each contour. The vertical dashed line
  shows the semimajor axis that is roughly equal to 2 stellar radii,
  assuming $R_\star = 4$~\rsun, which is typical of stars in the Keck
  survey. The position of \starA\,b is marked with a solid
  circle. \label{fig:complete}}  
\end{figure}

Repeating this procedure for all 137 stars provides a 
measure the completeness of our survey for planets above a given
\msini\ at each semimajor axis sampled in our simulations.
Figure~\ref{fig:complete} shows contours of constant completeness for
our sample, together with the position of \starA, 
demonstrating that we are 95\% complete for \msini~$ > 1$~\mjup\ and $a
< 0.1$~AU. 

Since each detection or nondetection of a hot Jupiter represents a
Bernoulli trial, the fraction of stars with planets in our sampled
range is given by the binomial distribution P$(f|k,N) \propto
f^k (1-f)^{N-k}$, where $N=137$ is the number of target stars
containing $k$ detections. For our sample we measure
an occurrence rate of $f = 1.2^{+1.2}_{-0.7}$, or $f < 0.034$ at 95\%
confidence for \msini~$ \geq 0.9$~\mjup, which includes a single
detection, \starA\,b with \msini~$
= \msiniA \pm \msinieA$. Restricting our analysis to planets more
massive than 1~\mjup\ we find $f < 0.025$  at 95\% confidence. 

Based on their survey of main-sequence A-type stars,
\citet{lagrange09a} reported no planets with $P < 10$~days among a
sample of 50 stars\footnote{\citet{lagrange09a} do not report stellar
masses, but instead classify stars based on $B-V$ colors. We
selected stars from their Table 4 with colors $0.0 < B-V < 0.3$ as
representative of main-sequence A-stars in our mass range ($M_\star >
1.45$~\msun).}, or $f < 
0.05$ at 95\% confidence. However, because their early-type stars are
such rapid rotators, their achievable velocity precision only allowed
them to rule out planets with \msini~$ \gtrsim 8$~\mjup. Thus, our
estimate of the fraction of massive stars with hot Jupiters represents
a significant refinement compared to the previous
constraints. However, our sample size is still too small to provide a
meaningful comparison with the planet fraction measured by
\citet{cumming08} for less massive stars. An extension of our
subgiants survey to the Southern sky is therefore warranted, and is
currently underway at the Anglo Australian Observatory (R. Wittenmyer,
private communication). Additional constraints will be provided by the
\emph{Kepler} space-based transit survey \citep{kepler}. 

\section{Summary}

We report the discovery of a hot
Jupiter ($e = \eA \pm \eeA$, \msini~$ = \msiniA$~\mjup, $P =
\pA$~days) orbiting the 
subgiant \starA. At $M_\star = \mstarA$~\msun, \starA\ is the most
massive star known to harbor a 
hot Jupiter, and this short-period system is the first detected as
part of a Doppler survey of evolved, intermediate-mass stars. 

The existence of this planet demonstrates that the observed population
of close-in planets around subgiants is largely representative of the
``primordial'' planet population, in that close-in planets have not
been adversely affected by the relatively mild, post-main-sequence
expansion of their host stars. This aspect of subgiants, together with
their jitter-limited RV precision of $\approx 5$~\ms, makes them
ideally suited for studying the properties of short-period planets
with a wide range of minimum masses around intermediate-mass stars. 

Based
on our current stellar sample we estimate that
0.5--2.3\% of A-type stars harbor a planet with $a < 0.1$~AU and
\msini~$ > 1$~\mjup, compared to the 0.4\% occurrence rate around
Sun-like stars \citep{cumming08}. While planets with $a < 1$~AU are
unusually rare around A 
stars, it is possible that there exists a population of hot Jupiters
($a < 0.1$~AU) around intermediate-mass stars comparable to 
what is found around Sun-like stars. If so, then the close-in desert
around A stars may simply be an exaggerated version of the ``period
valley'' 
observed around Sun-like stars, marked by a deficit of planets with
periods ranging from roughly 10--100 days, a sharp increase in the
number of detected planets beyond 1~AU, and a pile-up near $P=3$~day
\citep{udrysantos07, cumming08, wright09}. Doppler surveys of a 
larger number of massive subgiants, together with careful analyses of
detections from transit surveys, will test this possibility.

\acknowledgments
We gratefully acknowledge the efforts and dedication
of the Keck Observatory staff, especially Grant Hill, Scott Dahm and
Hien Tran for their support of 
HIRES and Greg Wirth for support of remote observing. We are also
grateful to the time assignment committees of NASA, NOAO, Caltech, and
the University of California for their generous allocations of
observing time. A.\,W.\,H.\ gratefully acknowledges support from a
Townes Post-doctoral Fellowship at the U.\,C.\ Berkeley Space Sciences 
Laboratory. G.\,W.\,M.\ acknowledges NASA grant NNX06AH52G.   
G.\,W.\,H acknowledges support from NASA, NSF,
Tennessee State University, and the State of Tennessee through its
Centers of Excellence program. 
Finally, the authors wish to extend special thanks to those of
Hawaiian ancestry  on whose sacred mountain of Mauna Kea we are
privileged to be guests.   
Without their generous hospitality, the Keck observations presented herein
would not have been possible.

\bibliography{}

\end{document}